\documentclass[12pt]{article}
\usepackage{cite,graphicx,amsmath,amsfonts,amssymb,mathrsfs}

\newcommand{\ie}{{\it i.e.}}

\newcommand{\eg}{{\it e.g.}}

\newcommand{\cf}{{\it cf.}}

\newcommand{\eq}{Eq.}

\newcommand{\fig}{Fig.}

\newcommand{\Ref}{Ref.}
\newcommand{\Refs}{Refs.}
\newcommand{\Sec}{Sec.}

\begin{document}

\title{
\vspace*{-3cm}
\begin{flushright}
{\small TUM-HEP-422/01}
\end{flushright}
{\bf Peculiar effects in the combination of neutrino decay and neutrino
oscillations} \\ \vspace*{0.5cm} {\normalsize
Talk given at the \\ ESF-NORDITA WORKSHOP ON NEUTRINO PHYSICS AND COSMOLOGY \\
June 11-22, 2001, Copenhagen, Denmark} } 

\author{Walter Winter\thanks{E-mail: {\tt wwinter@physik.tu-muenchen.de}} \\
\small{Institut f{\"u}r Theoretische Physik, Physik-Department} \\
\small{Technische Universit{\"a}t M{\"u}nchen} \\
\small{James-Franck-Stra{\ss}e, 85748 Garching bei M{\"u}nchen, Germany} }

\date{}

\maketitle

\begin{abstract}
In this talk, we will demonstrate some concepts of a
simultaneous treatment of neutrino decays and neutrino
oscillations in an illustrative manner. This includes topics such as phase coherence discussions
and time delay effects of massive supernova neutrinos.
\end{abstract}

\section{Introduction}

Neutrino decay in vacuum has often been considered as an alternative
to neutrino oscillations (\eg, in
\cite{Pakvasa:1972gz,Bahcall:1972my,Raghavan:1988ue,Acker:1992eh,Barger:1998xk,Barger:1999bg,Fogli:1999qt,Choubey:2000an,Bandyopadhyay:2001ct}).
Either neutrino decay only (especially for atmospheric
or solar neutrinos) or sequential combinations of neutrino
oscillations and decays (especially neutrino oscillations in matter followed by neutrino
decay in vacuum: MSW-mediated/MSW-catalyzed solar
neutrino decay) have been studied. However, simultaneous neutrino
decays and oscillations are also a possible scenario
\cite{Lindner:2001fx}. It involves several quantum field
theoretical issues such as phase
coherence \cite{Lindner:2001fx,Lindner:2001th}. In this talk, we
will  show some peculiarities coming from this sort of discussions,
introduced in a quite conceptual manner and illustrated by several examples.

\section{Majoron decay as an example}

In order to demonstrate several kinematics and coherence issues of a
decay model, we choose Majoron decay as an example
\cite{Zatsepin:1978iy,Chikashige:1980qk,Gelmini:1981re,Pakvasa:1999ta}. Let us
assume a generic effective interaction Lagrangian such as
\begin{equation} \label{int2}
\mathcal{L}_{\mathrm{int}} = \underset{i \neq j}{\sum_{i} \sum_{j}} g_{ij}
\overline{\nu_{j,L}^{c}} \nu_{i,L} J,
\end{equation}
where $J$ is the Majoron field, $\nu_i$ are Majorona mass eigenstates, and
$g_{ij}$ are the Majoron coupling constants. First of all, we observe
that only mass eigenstates and not flavor eigenstates may decay. Second, decay
into active as well as sterile neutrinos is, in principle, possible with this
type of Lagrangian. Third, we assume the secondary active neutrinos to be, in
principle, observable, but the Majorons, to first order, not.

\subsection{Re-direction of neutrinos by decay}

Re-direction of neutrinos by decay is a purely kinematical effect.
Since (at least) a third particle is involved in a decay process, the
secondary neutrino may slightly change direction because of energy and
momentum conservation. Let us now investigate the consequences for neutrino beams
and radially symmetric point sources.

\subsubsection*{Neutrino beams}

\begin{figure}[ht!]
\begin{center}
\includegraphics*[height=3cm]{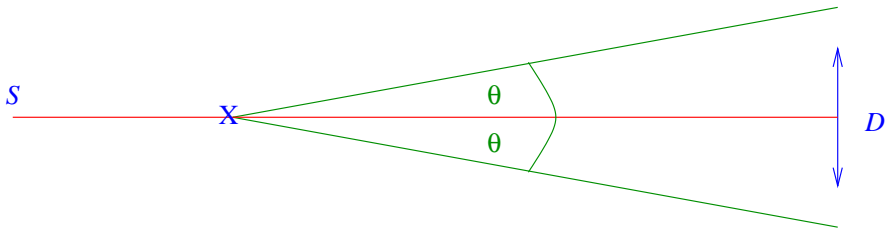}
\end{center}
\caption{\label{Beam} A neutrino beam produced by the source $S$ and detected
at the detector $D$, where $D$ is also the area of detection. The decay may
happen at $X$, changing the direction of the secondary neutrino by an angle
$\theta$.} \end{figure}

Figure~\ref{Beam} shows the geometry of a neutrino beam produced at
$S$ and directed towards
the detector $D$ with an intermediate decay at $X$.
One can derive from the kinematics of Majoron decay that the angle $\theta$
is limited by a maximum angle $\theta_{\mathrm{max}}$ for $\nu_i \rightarrow
\nu_j$ decay
\begin{equation}
\theta_{\mathrm{max}} = \frac{m_j}{2 E_i} \frac{\Delta
m_{ij}^2}{m_j^2},  \quad  \Delta m_{ij}^2 \equiv m_i^2-m_j^2>0.
\end{equation}
The angle $\theta_{\mathrm{max}}$ is determined by the
$\frac{m_j}{E_i}$-dependence for a not too hierarchical mass
spectrum. Thus, we obtain for relativistic neutrinos $\theta_{\max}
\ll 1$. One can show that
active secondary neutrinos are, in principle, observable for accelerator,
atmospheric, and reactor neutrinos ($\Delta m_{ij}^2 \simeq m_j^2$
assumed) \cite{Lindner:2001fx}.

\subsubsection*{Radially symmetric neutrino sources}

\begin{figure}[ht!]
\begin{center}
\includegraphics*[height=6.5cm]{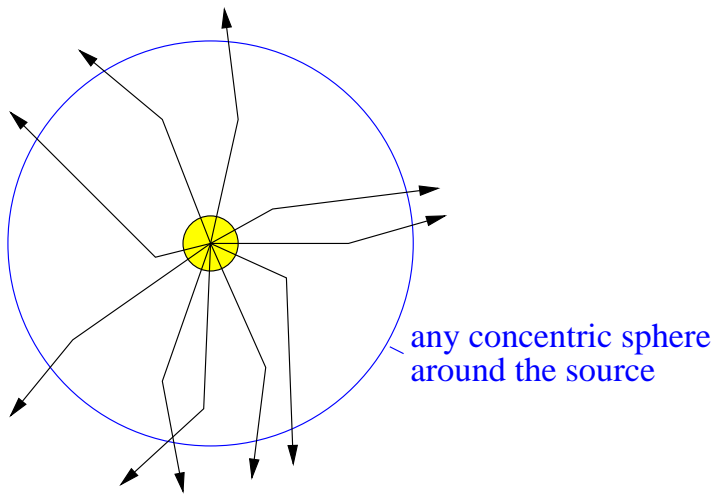}
\end{center}
\caption{\label{point} A radially symmetric neutrino source and
several paths of neutrinos with one intermediate decay. Drawing any concentric
sphere around the source illustrates that the overall flux through any
of these spheres is conserved as well as radially symmetric.}
\end{figure}

From \fig~\ref{point} we see that for decay
of one parent neutrino into exactly one secondary neutrino the overall flux of a radially
symmetric neutrino source is conserved.
Thus, active secondary neutrinos are, in principle, observable for solar
and supernova neutrinos. Furthermore, we notice that especially for
massive supernova
neutrinos, the travel times on different paths may be different even for
small angles $\theta_{\mathrm{max}}$. As we will see later, this will modify
the time dependence of the signal at the detector.

\subsection{Interference effects}

In this section, we will study phase coherence in
decay processes. Especially, we are interested in the observability of
interference effects with intermediate decays between production and
detection. Neutrino oscillations of decay products will be one example
for such an interference effect.

Neglecting the Majoron field as well as the operators, we know that to
first order in the S-matrix expansion
\begin{equation}
 d \Gamma \propto \left|  \langle out | \int\limits_{-\infty}^{\infty} d^4 x
\mathcal{L}_{\mathrm{int}} | in 
\rangle  \right|^2, \quad \mathcal{L}_{\mathrm{int}} \sim
\sum\limits_{ij} g_{ij} \bar{\nu}_j \nu_i.
\label{dGamma}
\end{equation}
Thus, the interaction destroys an $in$ state and creates an $out$ state by
application of the appropriate annihilation and creation operators in the
field expansions within the Lagrangian. Let us now assume an incoming
and outgoing superposition of
mass eigenstates, \ie, active flavor eigenstates (ignoring
neutrino propagation):
\begin{eqnarray}
 | in \rangle & = & | \nu_{\alpha} \rangle  = \sum\limits_i U_{\alpha i}^* |
\nu_i \rangle, \\ 
 | out \rangle & = &  | \nu_{\beta} \rangle = \sum\limits_j U_{\beta j}^* |
\nu_j \rangle.
\end{eqnarray}
Applying this to \eq~(\ref{dGamma}) shows that the
differential decay rate may indeed contain interference terms $\propto \langle
\nu_j | S_{ij} | \nu_i \rangle^* \langle \nu_l | S_{kl} | \nu_k
\rangle$ with $(i,j) \neq (k,l)$, corresponding to interference of different decay channels.

\begin{figure}[ht!]
\begin{center}
\begin{minipage}[h!]{13cm}
\begin{minipage}[t!]{6cm}
\includegraphics*[width=5.8cm]{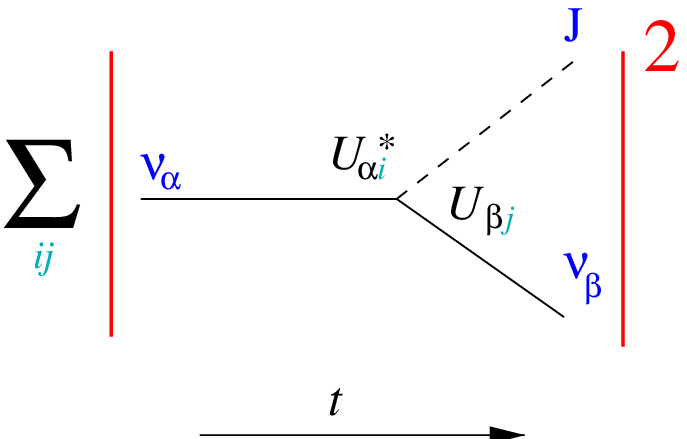}
\end{minipage}
\hspace*{1cm}
\begin{minipage}[t!]{6cm}
\includegraphics*[width=5.8cm]{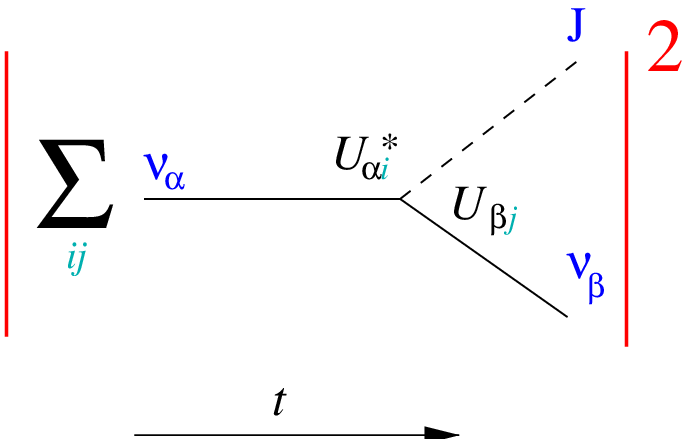}
\end{minipage}
\end{minipage}
\end{center}
\caption{\label{cohsum} Incoherent (left) and coherent
(right) summation of amplitudes for the case of Majoron decay.}
\end{figure}

\begin{figure}[ht!]
\begin{center}
\begin{minipage}[h!]{13cm}
\begin{minipage}[t!]{6cm}
\includegraphics*[width=5.8cm]{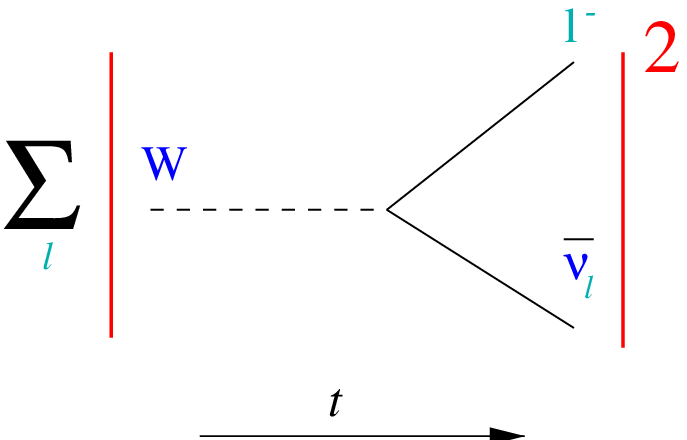}
\end{minipage}
\hspace*{1cm}
\begin{minipage}[t!]{6cm}
\includegraphics*[width=5.8cm]{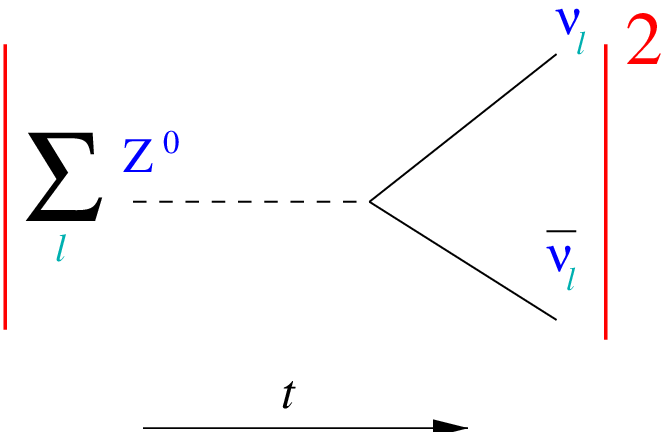}
\end{minipage}
\end{minipage}
\end{center}
\caption{\label{cohsumweak} Incoherent (left) and coherent
(right) summation of amplitudes for some cases of $W$ (left) and
$Z^0$ (right) decay.}
\end{figure}

From a different point of view, interference of different decay channels
corresponds to coherent summation of amplitudes. This is illustrated in
\fig~\ref{cohsum} for the case of Majoron decay.
Let us compare this to the examples of weak interaction processes shown in
\fig~\ref{cohsumweak}.
In the case of $W$ decay, the coherence among the $out$ neutrinos of different
flavors is destroyed by the mass differences of the participating leptons
corresponding to the different flavors. In other words, the wave
packets of the different neutrino flavors do not sufficiently overlap
because of the kinematics of the different leptons.
Thus, the Feynman diagrams for different flavors need to be summed
incoherently. However, in the case of $Z^0$ decay, the produced neutrino-antineutrino
pairs have small enough mass differences to allow wave
packet overlaps of different flavor or mass eigenstates. This means that we
need to sum the Feynman diagrams coherently \cite{Smirnov:1992eg}.

\begin{figure}[ht!]
\begin{center}
\begin{minipage}[h!]{13cm}
\begin{minipage}[t!]{6cm}
\includegraphics*[width=5.8cm]{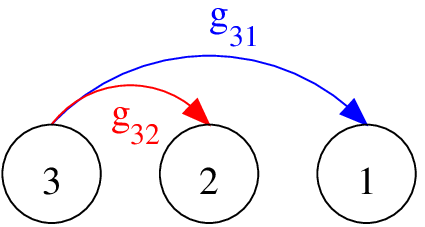}
\end{minipage}
\hspace*{1cm}
\begin{minipage}[t!]{6cm}
\includegraphics*[width=5.8cm]{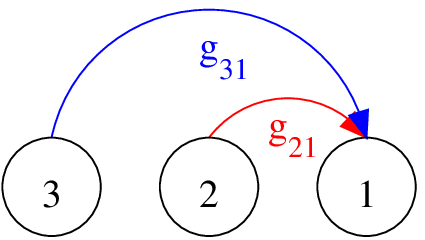}
\end{minipage}
\end{minipage}
\end{center}
\caption{\label{interf} Examples of decay scenarios in which neutrino
oscillations of secondary neutrinos (left) or neutrino oscillations of
parent neutrinos (right) become, in principle, possible. For these scenarios the Majoron
coupling constants shown in the figure need to have non-zero values. The
circles correspond to the mass eigenstates $\nu_3$, $\nu_2$, and
$\nu_1$, respectively.}
\end{figure}

Comparing $Z^0$ decay to Majoron decay, we conclude that interference is not
necessarily destroyed in neutrino decay. Thus, depending on the decay
scenario, we may have to take into account neutrino oscillations of
secondary neutrinos or parent neutrinos, such as illustrated in \fig~\ref{interf}.

\subsection{Decay as measurement}

\begin{figure}[ht!]
 \begin{center}
\includegraphics*[height=4cm]{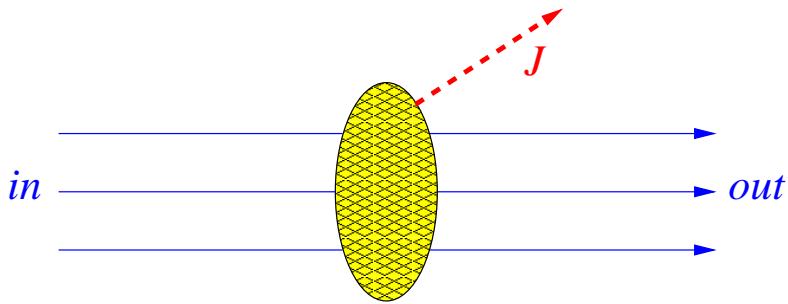}
\end{center}
\caption{\label{decayproc} Decay process illustrated, with an $in$
and $out$ superposition of mass eigenstates.}
\end{figure}

\begin{figure}[ht!]
\begin{center}
\begin{minipage}[h!]{13cm}
\begin{minipage}[t!]{6cm}
\includegraphics*[width=5.8cm]{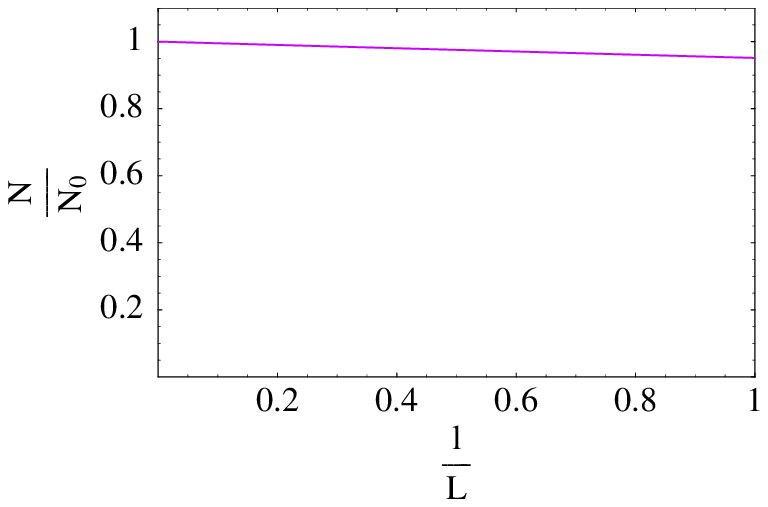}
\end{minipage}
\hspace*{1cm}
\begin{minipage}[t!]{6cm}
\includegraphics*[width=5.8cm]{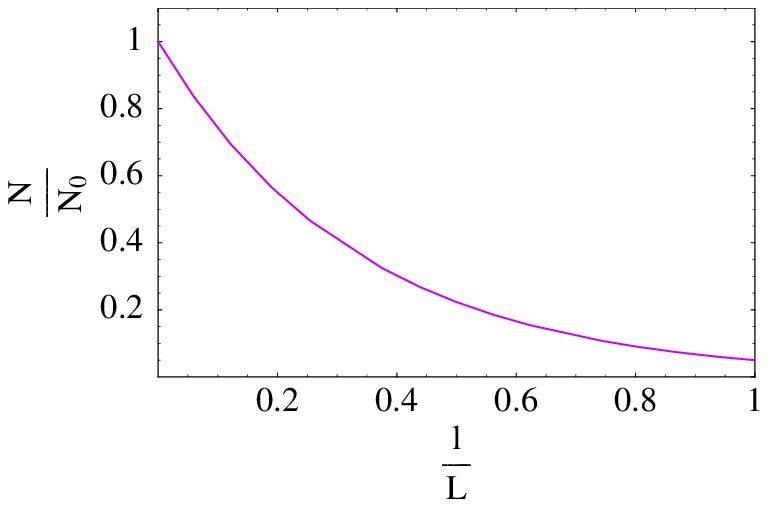}
\end{minipage}
\end{minipage}
\end{center}
\caption{\label{decayrates} Fraction of undecayed neutrinos $N/N_0$ over the
fraction of the baseline $l/L$ for small (left) and large (right)
decay rates (neutrinos treated as particles).
For exponential decay $\propto \exp(-\alpha l/E)$ this corresponds to $\alpha
\ll E/L$ (left) and $\alpha \sim E/L$ (right). }
\end{figure}

Similar to the production and detection processes in neutrino oscillations,
neutrino decay acts as a measurement. In general, decay destroys
an incoming superposition of mass eigenstates and creates an outgoing
one, as it is indicated in \fig~\ref{decayproc}.
If we can only detect the secondary neutrinos but not the Majorons, we
will in many cases (for unchanged quantum numbers and similar energies) not
even be able to tell, if there has been a decay between production and
detection, or not. However, since there is a third particle involved (the
Majoron), these two cases can, in principle, be
distinguished. This is equivalent to the fact that Feynman diagrams of
different orders do not interfere. Thus, decay acts as a measurement and
resets the relative phase among the mass eigenstates in the $out$ state to
$0$, similar to the example of $Z^0$ decay above. 
Therefore, for any considered neutrino oscillation (before or after decay)
the oscillation phases at the detector depend on the position of the decay.
Figure~\ref{decayrates} shows the number of neutrinos over the traveling
distance for small and large decay rates.
For small decay rates, the positions
of decay are almost equally spread over the entire traveling distance. Thus, any
oscillation phase will be averaged out, similar to the case of production or detection
regions larger than the oscillation length. However, for large
decay rates more neutrinos will decay in the beginning of the
path than at the end. Since the oscillation phases are averaged over
all possible decay positions, we may thus expect a net oscillatory
effect.

\section{Invisible decay products}

For decay into unobservable particles, such as sterile decoupled neutrinos,
we do not have to take care of secondary neutrinos as well as the type of
neutrino source. Therefore, this is the simplest case of neutrino
decay. One can show that the transition probability is given by
\cite{Lindner:2001fx} \begin{eqnarray}
P_{\alpha \beta}^{\rm invisible} & = & 
\underbrace{\sum\limits_{ij} \Re J_{ij}^{\alpha \beta} e^{-
\Gamma_{ij}} - 4 \; \underset{i>j}{\sum\limits_{ij}} 
\Re J_{ij}^{\alpha \beta} \sin^2 \Delta_{ij} e^{-\Gamma_{ij}}}_{P_{\rm
CP \, conserving}} \nonumber \\ & - & \underbrace{ 2 \;
\underset{i>j}{\sum\limits_{ij}} \Im J_{ij}^{\alpha \beta} \sin 2
\Delta_{ij} e^{- \Gamma_{ij}}}_{P_{\rm CP \, violating}}
\label{transprobdecay}
\end{eqnarray}
with
\begin{eqnarray}
J_{ij}^{\alpha \beta} & \equiv & U_{\alpha i} U^{*}_{\alpha j} U^{*}_{\beta i}
U_{\beta j}, \quad  \Delta_{ij} \equiv \frac{\Delta m^2_{ij} L}{4 E},
\\
 \Gamma_{ij} & \equiv &  \left( \frac{m_{i}
}{\tau_{0,i}} + \frac{m_{j}}{\tau_{0,j}} \right) \frac{L}{2 E} = \left(
\alpha_{i}+\alpha_{j} \right) \frac{L}{2 E}.
\end{eqnarray}
Here $\alpha_{i} \equiv m_{i}/\tau_{0,i}$, where
$\tau_{0,i}$ is the (rest frame) lifetime of $\nu_i$. In
\eq~(\ref{transprobdecay}) we see that the oscillatory terms are damped by
exponentials describing the disappearance of neutrinos into decay products
invisible to the detector.

\subsubsection*{Example: Atmospheric neutrino decay}

\begin{figure}[ht!]
\begin{center}
\includegraphics*[width=5.8cm]{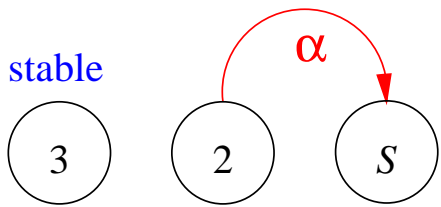}
\end{center}
\caption{\label{atmmodel} A scenario for atmospheric neutrino decay. Here
the mass eigenstate $\nu_2$ may decay into a sterile eigenstate, which is not
mixing with the active mass eigenstates.}
\end{figure}

\begin{figure}[ht!]
\begin{center}
\includegraphics*[height=9cm]{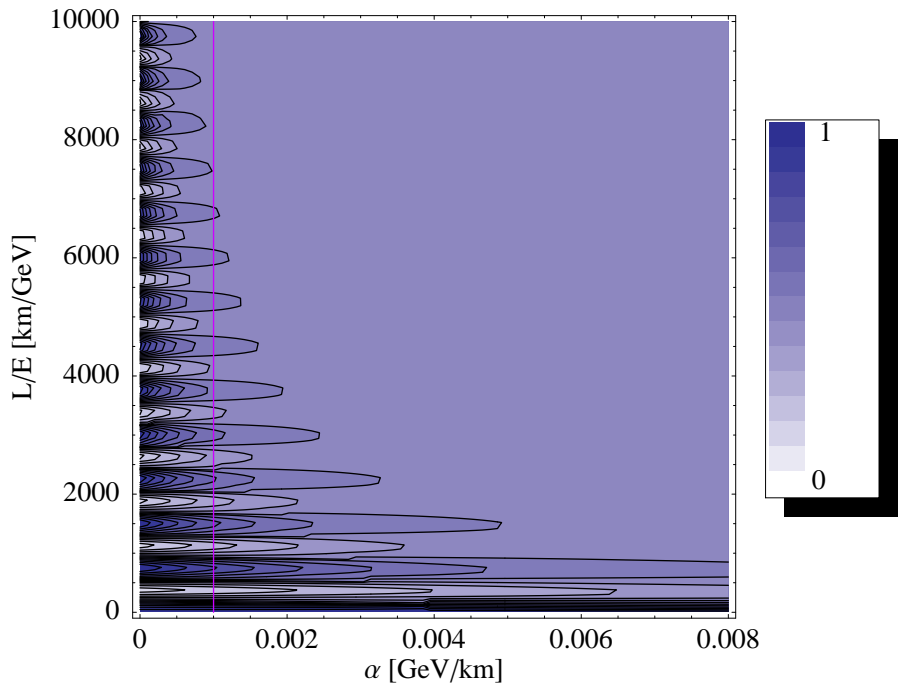}
\end{center}
\caption{\label{DecayOsc} Contour plot of the survival probability $P_{\mu
\mu}$ over the decay rate $\alpha$ and the sensitivity $L/E$ (similar plot
as in \Ref \cite{Lindner:2001fx}). The decay scenario used is shown in
\fig~\ref{atmmodel}. For the atmospheric neutrino oscillation parameters we
take $\cos^2 \theta_{23} = 0.30$ and $\Delta m_{32}^2 = 3.3 \cdot 10^{-3} \,
\mathrm{eV}^2$.} \end{figure}

\begin{figure}[ht!]
\begin{center}
\includegraphics*[height=7cm]{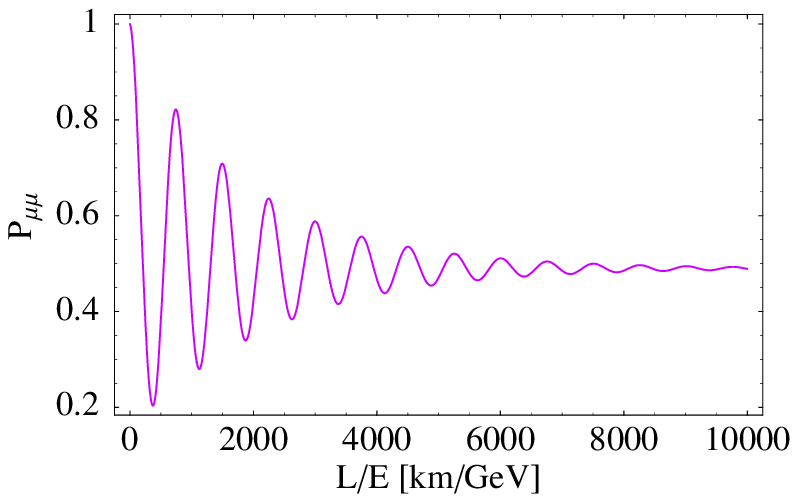}
\end{center}
\caption{\label{DecayOscCut}
 Cut through \fig~\ref{DecayOsc} at $\alpha = 1/1000 \,
 \mathrm{GeV/km}$ at the purple line.}
\end{figure}

In order to demonstrate the effects of invisible decay, we may choose a
decay scenario similar to the one in \Ref~\cite{Barger:1999bg} shown in
\fig~\ref{atmmodel} for atmospheric neutrino decay.
For the parameters we take $\cos^2
\theta_{23} = 0.30$ \cite{Barger:1999bg} as well as $\Delta m_{32}^2 = 3.3
\cdot 10^{-3} \, \mathrm{eV}^2$, since we want to take into
account neutrino oscillations in addition to neutrino
decay. Figure~\ref{DecayOsc} shows the survival
probability $P_{\mu \mu}$ for different decay rates $\alpha$ and sensitivities
$L/E$.
One can imagine the transition curve between decay and oscillation
dominated regions, determined by $\alpha L/E \sim 1$. The purple line
in \fig~\ref{DecayOsc} corresponds to the cut shown in
\fig~\ref{DecayOscCut}, in which the damping of the oscillation by
decay can be clearly seen.

\section{Visible decay products in neutrino beams}

We showed for the example of Majoron decay that decay into active neutrinos,
or sterile neutrinos mixing with active ones, involves more
complicated discussions than decay into invisible neutrinos, such as
sterile decoupled neutrinos. In this section,
we will only give a notion of the results for visible secondary
neutrinos in neutrino beams.

Let us define $P_{\alpha \beta}^{i}$ to be the transition probabilities for 
the flavor transition $\nu_{\alpha} \rightarrow \nu_{\beta}$ with {\em exactly}
$i$ intermediate decays ($i=0,1,2,\hdots$). In order to calculate the total
transition probability, the transition probabilities for different indices
$i$ have to be summed over or not. This depends on the ability to
distinguish the secondary from the parent neutrinos, \ie, conceptual
properties of the detector and the problem. For example, for decay into
antiparticles the decay products and the parent neutrinos may have different
signatures in the detector. Another example is energy resolution: since
neutrinos loose some energy to third particles by decay, the detector may
distinguish the parent and secondary neutrinos by its energy resolution.

Assuming that no secondary neutrinos escape detection by kinematics, one can
show for the first transition terms $P_{\alpha \beta}^{i}$ that \cite{Lindner:2001fx}
\begin{eqnarray}
P_{\alpha \beta}^{0} & \equiv & P_{\alpha \beta}^{\rm invisible}, \\
P_{\alpha \beta}^{1} & = & \underset{i \neq
j}{\sum\limits_{ij}} \, \underset{k \neq l}{\sum\limits_{kl}} {L \over E}
\frac{\sqrt{\alpha_{ij} \alpha_{kl}}}{\left( \Gamma_{jl}-\Gamma_{ik} \right)^2
+ 4 \left( \Delta_{ij} + \Delta_{lk} \right)^2 } \nonumber \\ &\times& \bigg\{
\Re(K_{ijkl}^{\alpha \beta}) \left[ \left( \Gamma_{jl} - \Gamma_{ik} \right)
\left( e^{- \Gamma_{ik}} \cos ( 2 \Delta_{ki} ) - e^{-\Gamma_{jl}} \cos ( 2
\Delta_{lj} ) \right) \right. \nonumber \\ &-& \left. 2 \left( \Delta_{ij} +
\Delta_{lk} \right) \left( e^{- \Gamma_{ik}} \sin ( 2 \Delta_{ki} ) -
e^{-\Gamma_{jl}} \sin ( 2 \Delta_{lj} ) \right) \right] \nonumber \\
&-& \Im(K_{ijkl}^{\alpha \beta}) \left[ \left( \Gamma_{jl} -
\Gamma_{ik} \right) \left( e^{- \Gamma_{ik}} \sin ( 2 \Delta_{ki} ) -
e^{-\Gamma_{jl}} \sin ( 2 \Delta_{lj} ) \right) \right. \nonumber \\
&+& \left. 2 \left( \Delta_{ij} + \Delta_{lk} \right) \left( e^{-
\Gamma_{ik}} \cos ( 2 \Delta_{ki} ) - e^{-\Gamma_{jl}} \cos ( 2 \Delta_{lj} )
\right) \right] \bigg\} 
\label{pone}
  \\
P_{\alpha \beta}^{2} & = &  \hdots, \nonumber
\end{eqnarray}
where $P_{\alpha \beta}^{\rm invisible}$ is given by
\eq~(\ref{transprobdecay}), $K^{\alpha \beta}_{ijkl} \equiv U^*_{\alpha i}
U_{\beta j} U_{\alpha k} U_{\beta l}^*$ is a generalization of $J^{\alpha
\beta}_{ij}$, and $\alpha_{ij} \equiv m_{i}/\tau_{0,ij}$ is the decay rate for
the channel $\nu_i \rightarrow \nu_j$, analogously defined to $\alpha_{i}
\equiv m_{i}/\tau_{0,i} \equiv \sum_j \alpha_{ij}$.

\subsubsection*{Example}

\begin{figure}[ht!]
\begin{center}
\includegraphics*[width=5.8cm]{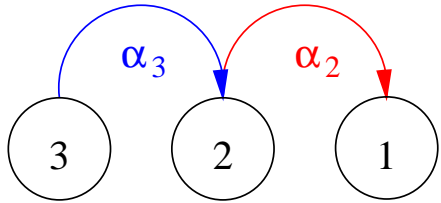}
\end{center}
\caption{\label{oscmodel} Example for a decay scenario with $m_3>m_2>m_1$.}
\end{figure}

\begin{figure}[ht!]
\begin{center}
\includegraphics*[height=7cm]{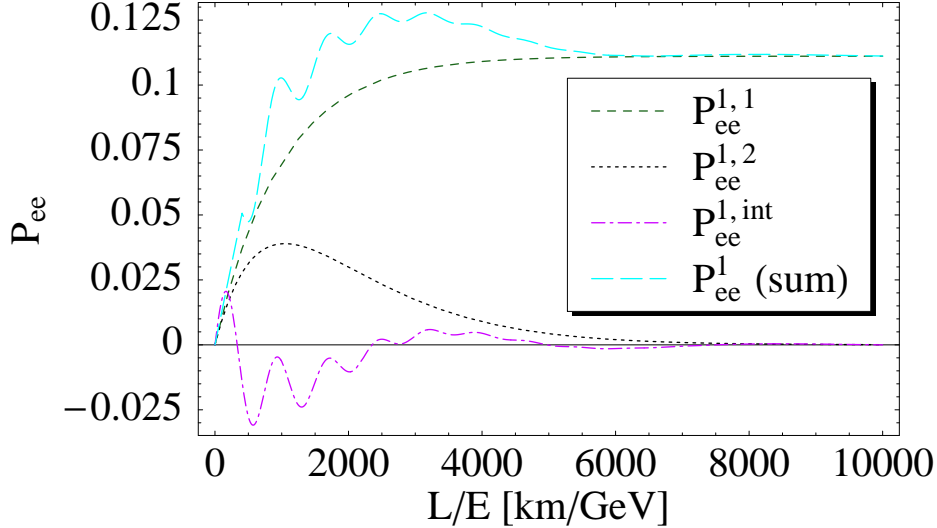}
\end{center}
\caption{\label{appsep} The separated survival probabilities
$P_{ee}^{1,k}$ for decay into individual mass eigenstates $\nu_k$, the
interference probability $P_{ee}^{1,\mathrm{int}}$, as well as the total
survival probability with exactly one intermediate decay
$P_{ee}^{1}$ (similar plot as in \Ref \cite{Lindner:2001fx}). Here the decay
scenario in \fig~\ref{oscmodel} is used with parameter values $\Delta
m_{32}^2 = 3.3 \cdot 10^{-3} \, \mathrm{eV}^2$, $\Delta m_{21}^2 = 5 \cdot
10^{-4} \, \mathrm{eV}^2$, $\alpha_2 = 1/1000 \, \mathrm{GeV/km}$, $\alpha_3 =
1/1100 \, \mathrm{GeV/km}$, as well as trimaximal mixing.} \end{figure}

In order to show the effects for visible decay products, we construct an
example with the decay scenario shown in \fig~\ref{oscmodel}.
Let us look at the survival probability of electron neutrinos $P_{ee}$
and ignore higher order decay effects, \ie, only consider $P_{ee}^0$ and
$P_{ee}^1$. For the decay scenario in \fig~\ref{oscmodel} one can
split up the non-vanishing terms in the sum in $P_{ee}^1$ in \eq~(\ref{pone}) into
\begin{equation}
 P_{ee}^1 = P_{ee}^{1,1} + P_{ee}^{1,2} + P_{ee}^{1,\mathrm{int}}, 
 \label{psplit}
\end{equation}
where $P_{ee}^{1,k}$ describes the production of new $\nu_{k}$ by
decay and $P_{ee}^{1,\mathrm{int}}$ the interference effects
(neutrino oscillations) before or after decay. Figure~\ref{appsep} shows the
separated signals and their sum for the parameters in the figure caption.
 The
terms $P_{ee}^{1,k}$, describing the production of new mass eigenstates by
decay, are exponentially growing in the beginning. For large $L/E$,
$P_{ee}^{1,2}$ is falling again because $\nu_2$ decays with a larger rate than
being produced. The interference term $P_{ee}^{1,\mathrm{int}}$ is basically
determined by the two beat frequencies induced by the two $\Delta m^2$
involved, \ie, neutrino oscillations before and after decay.

\begin{figure}[ht!]
\begin{center}
\includegraphics*[height=7cm]{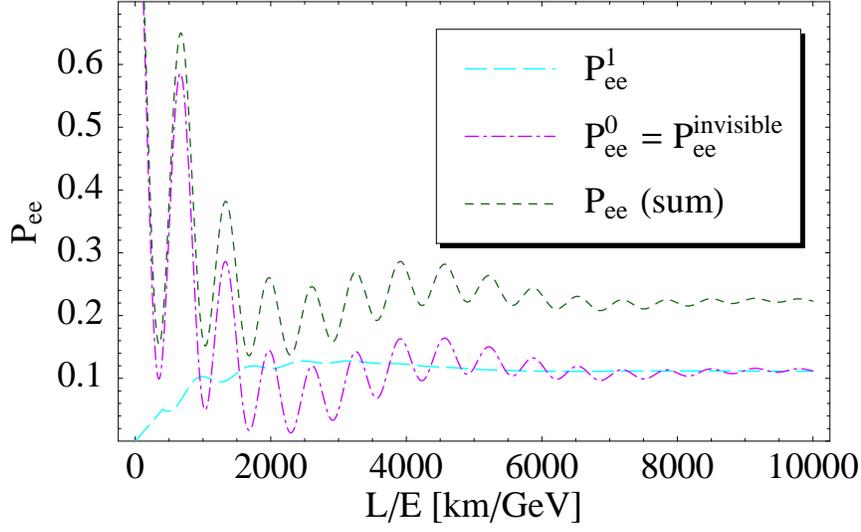}
\end{center}
\caption{\label{apptotal} The survival probabilities without
intermediate decays $P_{ee}^0$ and with one intermediate decay $P_{ee}^1$
between production and detection, as well as their sum (similar plot
as in \Ref \cite{Lindner:2001fx}).
The parameter values are chosen as given in the caption of \fig~\ref{appsep}.}
\end{figure}

Taking $P_{ee}^0$ into account, which describes the survival
probability of the neutrinos
arriving at the detector without any decay between production and detection,
yields the result in \fig~\ref{apptotal}.
Note that $P_{ee}^0$ and $P_{ee}^1$ may only be sensibly added, if the
detector cannot distinguish between parent and secondary neutrinos.

\section{Decays of supernova neutrinos}

In this section, we will focuse on supernova neutrinos. We will show
certain properties of supernova neutrino propagation as well as the effects
modifying neutrino event rates.

\subsection{Issues especially concerning supernova neutrinos}

Since a supernova may be approximated as a far-distant point
source, we have to incorporate some new concepts (\eg, \Refs
\cite{Aharonov:1988ju,Lindner:2001th}):

\subsubsection*{Time delays of massive neutrinos}

\begin{figure}[ht!]
\begin{center}
\includegraphics*[width=8cm]{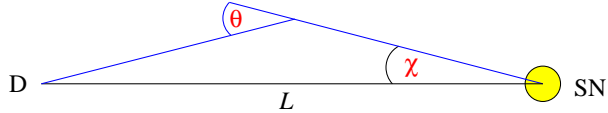}
\end{center}
\caption{\label{Geometry} 
A traveling path from the supernova $SN$ to the detector $D$ different
from the direct path along the baseline $L$.}
\end{figure}

For massive neutrinos the velocity of propagation depends on the mass. Even on
the direct path  $L$, massive neutrinos will be delayed by $\Delta t
\simeq L m^2/(2E^2)$. In addition, re-direction by decay opens the
possibility for paths from the supernova to the detector other than the direct
path, such as shown in \fig~\ref{Geometry}.
Again, neutrinos will be delayed by an additional time interval, though the
change of direction by decay is in many cases quite small.

In order to investigate this effect, let us assume a radially
symmetric source producing $N_{\alpha}$ neutrinos of flavor
$\nu_{\alpha}$ at $t=-L$, \ie,
\begin{equation}
 \Phi^{\mathrm{tot}}_{\alpha}(t) = \frac{dN_{\alpha}}{dt} =
N_{\alpha} \, \delta(t+L),
\end{equation}
so that
\begin{equation}
  \int\limits_{-\infty}^{\infty} \Phi^{\mathrm{tot}}_{\alpha}(t)
dt  =  N_{\alpha}.
\end{equation}
Therefore, for such a source flux massless neutrinos would arrive at $t=0$.
Let us further define $\Phi^{D,i}_{\alpha \beta}(t)$ to be the number of
neutrinos per time interval at the detector, which are produced as
flavor $\nu_{\alpha}$ and detected as flavor $\nu_{\beta}$ with exactly
$i$ intermediate decays. This definition is completely analogous to
$P^{i}_{\alpha \beta}$, but in addition takes into account the time dependence
of the signal.

\subsubsection*{Loss of coherence because of long baselines}

We know from the wave packet treatment of neutrino oscillations that the
coherence length of neutrino oscillations with $\Delta m_{ab}^2$ is given by
\cite{Giunti:1991ca,Giunti:1998wq,Grimus:1998uh,Cardall:1999ze}
\begin{equation}
L^{\mathrm{coh},I}_{ab} \equiv \frac{4 \sqrt{2} \sigma_x^I
E^2}{\Delta m_{ab}^2}.
\end{equation}
Here
\begin{equation}
 (\sigma_x^I)^2 = (\sigma_x^P)^2 + (\sigma_x^D)^2
\end{equation}
is the combined wave packet width $\sigma_x^I$ of the production $P$ and
detection $D$ processes. The decay process $X$ can be
treated similarly to an intermediate process between production $P$ and
detection $D$ by using
\begin{equation}
 (\sigma_x^I)^2 = (\sigma_x^P)^2 + (\sigma_x^X)^2 \quad \mathrm{or} \quad
(\sigma_x^I)^2 = (\sigma_x^X)^2 + (\sigma_x^D)^2,
\end{equation}
depending on what processes we are looking at.

\subsection{Dispersion by different neutrino masses}
\label{Sec:Disp}

\begin{figure}[ht!]
\begin{center}
\includegraphics*[width=5.8cm]{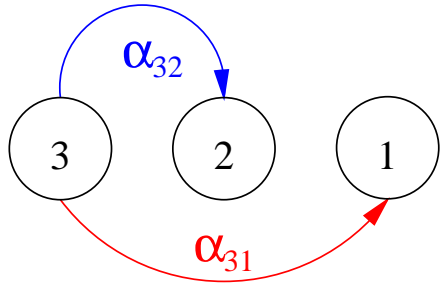}
\end{center}
\caption{\label{incohmodel} A decay scenario with
$m_3>m_2>m_1$ for decay of one mass eigenstate into two different decay channels.}
\end{figure}

\begin{figure}[ht!]
\begin{center}
\includegraphics*[height=7cm]{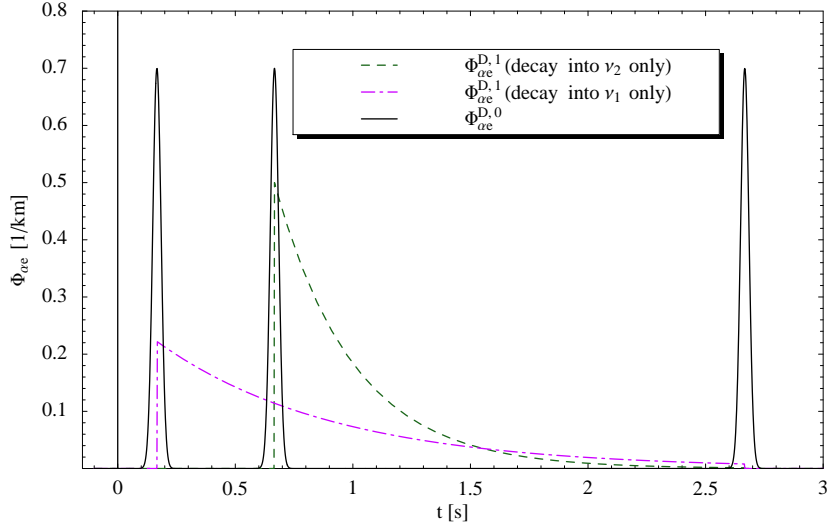}
\end{center}
\caption{\label{Dispersion} The signals for no intermediate decays
between production and detection $\Phi^{D,0}_{\alpha e}$, as well as
for one intermediate decay $\Phi^{D,1}_{\alpha e}$, separated into the terms for 
$\nu_1$ and $\nu_2$ as decay products (similar plot as in \Ref
\cite{Lindner:2001th}). In this example,
the decay scenario in \fig~\ref{incohmodel} with the parameter values
$m_3 = 4 \, \mathrm{eV}$, $m_2 = 2 \, \mathrm{eV}$, $m_1 = 1 \, \mathrm{eV}$,
$L=10^{22} \, \mathrm{m} \simeq 32 \, \mathrm{kpc}$, $E=10 \, \mathrm{MeV}$,
$\alpha_{32} = \alpha_{31} = E/L$, $N_{\alpha} = 9 \cdot 10^{5} \, 4 \pi
L^2/D$, as well as trimaximal mixing is used. Note that in this example 
the flavor index $\alpha$ may
 refer to any flavor. } \end{figure}

As a first effect, which we illustrate with an example, we demonstrate
the propagation of mass
eigenstates traveling with different velocities before and after decay
because of the different masses of parent and secondary neutrinos. Thus, the
arrival time depends on the position of decay. We choose
the decay scenario in \fig~\ref{incohmodel} and assume incoherent propagation at all times, \ie, the travel distance
between any two processes in the problem is much longer than the respective
coherence length.
In addition, we ignore effects of different traveling path
lengths as well as repeated decays. Figure~\ref{Dispersion}
shows the results for the parameter values given in the figure caption.
For no intermediate decays between production and detection,
the source pulses are also detected as pulses. For
one intermediate decay, the neutrinos travel with the velocity of the heavy
parent mass eigenstate before decay and with the velocity of the light
secondary mass eigenstate after decay. Thus, one can find the exponential
distribution of the decay positions in the time structure of the signal at
the detector. This is an effect which may affect or even mimic the signal structure
expected from supernova models.

\subsection{Dispersion by different traveling path lengths}

\begin{figure}[ht!]
\begin{center}
\includegraphics*[height=2.5cm]{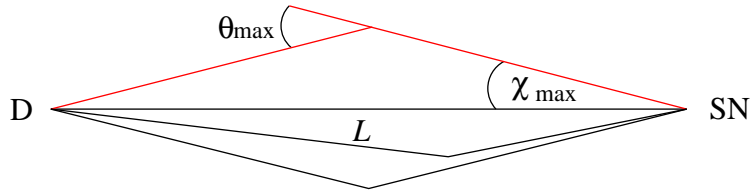}
\end{center}
\caption{\label{maxgeo} Different paths from the supernova $SN$ to the
detector $D$, as well as the path with maximum length by kinematics and
geometry (red).}
\end{figure}

\begin{figure}[ht!]
\begin{center}
\includegraphics*[width=5.8cm]{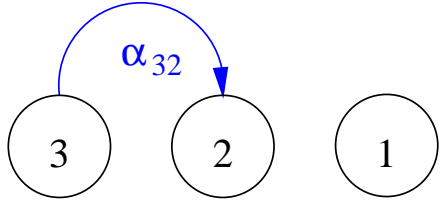}
\end{center}
\caption{\label{pathmodel} Example for a decay scenario with
$m_3>m_2>m_1$.}
\end{figure}

\begin{figure}[ht!]
 \begin{center}
\includegraphics*[height=7cm]{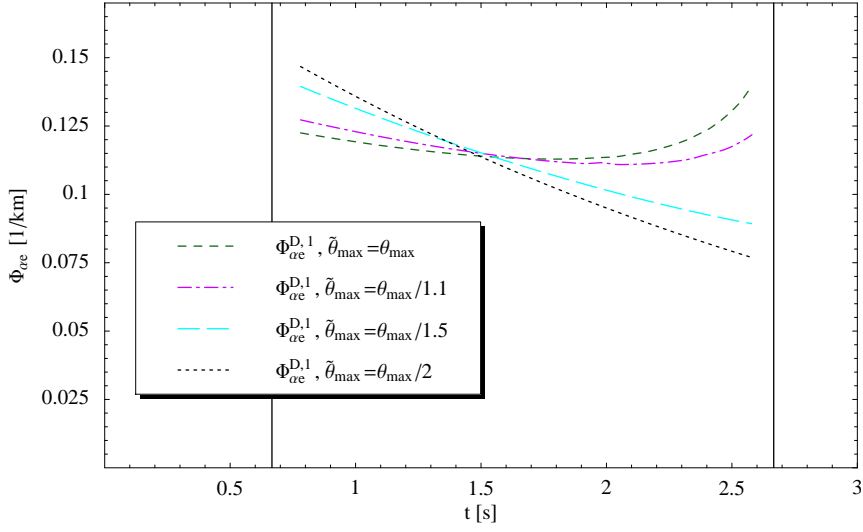}
\end{center}
\caption{\label{phidef2} Approximated signal with one intermediate decay for
several values of $\tilde{\theta}_{\mathrm{max}}$ defined in the
text (similar plot as in \Ref \cite{Lindner:2001th}). For the decay scenario we
use the one in \fig~\ref{pathmodel}. The parameter values are chosen such as
in the last example in the caption of \fig~\ref{Dispersion}. }
\end{figure}

In this section, we will illustrate the effects of different traveling path
lengths with an example. Note that the smaller the maximum
re-direction by decay $\theta_{\mathrm{max}}$ is, the smaller the
dispersion by time delays on different traveling paths becomes (\cf,
\fig~\ref{maxgeo}).
However, there will be dispersion by different neutrino masses even for small
$\theta_{\mathrm{max}}$, such as it was shown in the example
in \Sec~\ref{Sec:Disp}. We use an example with the decay
scenario shown in \fig~\ref{pathmodel} and the parameter values from the last
example in \Sec~\ref{Sec:Disp} (\cf, caption of \fig~\ref{Dispersion}).
In addition, we approximate the differential decay rate by its mean
\begin{equation}
  \frac{d \Gamma}{d \cos \theta} = \left\{
\begin{array}{cll} \frac{\Gamma_{\mathrm{tot}}}{1 - \cos
\tilde{\theta}_{\mathrm{max}}} & \mathrm{for} & \theta \le
\tilde{\theta}_{\mathrm{max}} \\
0 & \mathrm{otherwise} & \end{array} \right.
 \end{equation}
with an effective
$\tilde{\theta}_{\mathrm{max}} \le \theta_{\mathrm{max}}$. Thus, we may choose
$\tilde{\theta}_{\mathrm{max}}$ smaller than the actual
$\theta_{\mathrm{max}}$, in order to investigate the
dependence on the path lengths. Figure~\ref{phidef2} shows the (approximated)
signal with one intermediate decay for several values of
$\tilde{\theta}_{\mathrm{max}}$. 
The figure demonstrates that for large $\tilde{\theta}_{\mathrm{max}}$ late
time arrivals are favored and early time arrivals suppressed. For small
$\tilde{\theta}_{\mathrm{max}}$ we almost observe an exponential behavior
such as expected from the last example.

\subsection{Early coherent decays}

For this effect, we assume the decay rates to be large enough such that the neutrinos
are still coherently propagating at the position of decay, but loose
coherence between decay and detection. This also implies that all neutrinos
decay before detection.

\begin{figure}[ht!]
\begin{center}
\includegraphics*[width=5.8cm]{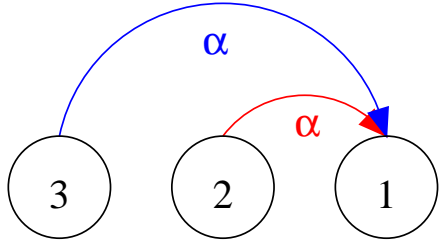}
\end{center}
\caption{\label{earlymodel} Example for a decay scenario with
$m_3>m_2>m_1$ and two decay channels into the same decay product.}
\end{figure}

\begin{figure}[ht!]
\begin{center}
\includegraphics*[height=7cm]{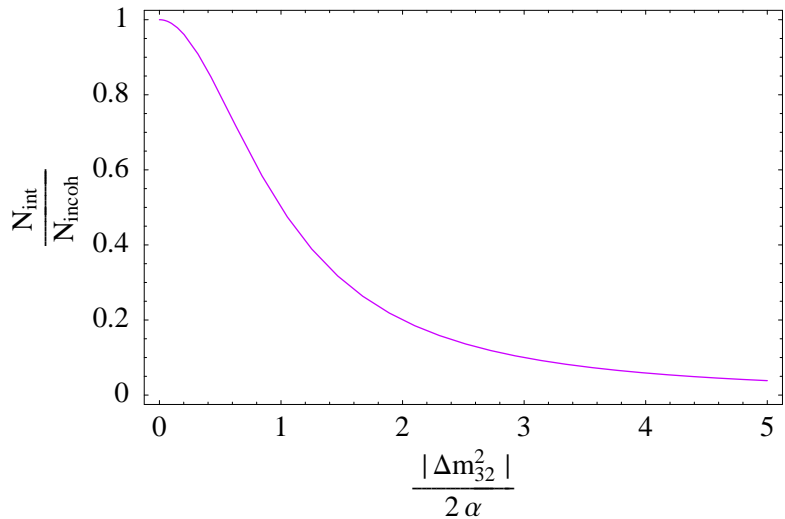}
\end{center}
\caption{\label{earlydecays} The ratio of neutrinos coming from the
calculation of interference terms $N_{\mathrm{int}}$ to the number of
neutrinos coming from incoherent propagation $N_{\mathrm{incoh}}$. This ratio
is plotted over the ratio of $\Delta m_{32}^2$, which causes
the interference terms, to the decay rate $\alpha$. For the calculation
trimaximal mixing as well as the decay scenario in \fig~\ref{earlymodel} is
assumed.}
\end{figure}

For the decay scenario we need to have simultaneous coupling of two mass
eigenstates to the decay product, such as in \fig~\ref{earlymodel}.
In addition, we assume trimaximal mixing. Since it can be shown that the
detector can in most cases not resolve the time dependence of the signal, we
integrate the flux over time: \begin{equation}
 N_{\alpha \beta}^{D,1} \equiv \int\limits_{-\infty}^{\infty} \Phi_{\alpha
\beta}^{D,1} dt.
\end{equation}
Moreover, similar to \eq~(\ref{psplit}), we define
$N_{\mathrm{int}}$ to be the number of neutrinos coming from interference
terms in the calculation and $N_{\mathrm{incoh}}$ to be the number of neutrinos
coming from incoherent propagation.
Figure~\ref{earlydecays} illustrates that for small $\Delta m_{32}^2 \ll
\alpha$ interference effects become most important.
For larger $\Delta m_{32}^2$ more and more oscillations take place
until the neutrinos decay and thus the oscillation phases at the position of
decay are more and more averaged out over all possible decay positions. Note
that we may observe an interference
effect, even if only one stable neutrino arrives at the detector.

\section{Summary}

So far, parameters have only been fitted for neutrino oscillations or
special neutrino decay scenarios. However, fitting all oscillation and decay
parameters may produce new solutions. Furthermore, decay effects for
supernova neutrinos have been ignored. Taking them into account may
alter or even mimic the signals expected from
supernova models. Finally, supernova neutrino observations have only indicated that at least one
mass eigenstate is stable. Nevertheless, interference phenomena may have to be
taken into account, even if only one stable mass eigenstate arrives at the
detector. Since non-zero values for neutrino masses imply, in principle, not
only neutrino oscillations, but also neutrino decay, we
conclude that neutrino decay should be incorporated into the general neutrino
oscillation discussion.

\subsubsection*{Acknowledgements}

The author would like to thank Manfred Lindner and Tommy Ohlsson for
useful discussions and comments, J{\"o}rn Kersten and Tommy Ohlsson for
proofreading the manuscript, and Lars Bergstr{\"o}m, Steen Hannestad,
Kimmo Kainulainen, and Georg Raffelt for organizing the workshop.

This work was supported by ESF (``European Science Foundation''), NORDITA,
the ``Studienstiftung des deutschen Volkes'' (German
National Merit Foundation), and the ``Sonderforschungsbereich 375
f{\"u}r Astro-Teilchenphysik der Deutschen Forschungsgemeinschaft''.


\begin{thebibliography}{10}

\bibitem{Pakvasa:1972gz}
S. Pakvasa and K. Tennakone,
\newblock Phys. Rev. Lett. 28 (1972) 1415.

\bibitem{Bahcall:1972my}
J.N. Bahcall, N. Cabibbo and A. Yahil,
\newblock Phys. Rev. Lett. 28 (1972) 316.

\bibitem{Raghavan:1988ue}
R.S. Raghavan, X.G. He and S. Pakvasa,
\newblock Phys. Rev. D38 (1988) 1317.

\bibitem{Acker:1992eh}
A. Acker, A. Joshipura and S. Pakvasa,
\newblock Phys. Lett. B285 (1992) 371.

\bibitem{Barger:1998xk}
V. Barger et~al.,
\newblock Phys. Rev. Lett. 82 (1999) 2640, astro-ph/9810121.

\bibitem{Barger:1999bg}
V. Barger et~al.,
\newblock Phys. Lett. B462 (1999) 109, hep-ph/9907421.

\bibitem{Fogli:1999qt}
G.L. Fogli et~al.,
\newblock Phys. Rev. D59 (1999) 117303, hep-ph/9902267.

\bibitem{Choubey:2000an}
S. Choubey, S. Goswami and D. Majumdar,
\newblock Phys. Lett. B484 (2000) 73, hep-ph/0004193.

\bibitem{Bandyopadhyay:2001ct}
A. Bandyopadhyay, S. Choubey and S. Goswami,
\newblock hep-ph/0101273.

\bibitem{Lindner:2001fx}
M. Lindner, T. Ohlsson and W. Winter,
\newblock Nucl. Phys. B  (to be published), hep-ph/0103170.

\bibitem{Lindner:2001th}
M. Lindner, T. Ohlsson and W. Winter,
\newblock (2001), astro-ph/0105309.

\bibitem{Zatsepin:1978iy}
G.T. Zatsepin and A.Y. Smirnov,
\newblock Yad. Fiz. 28 (1978) 1569,
\newblock [Sov. J. Nucl. Phys. 28 (1978) 807].

\bibitem{Chikashige:1980qk}
Y. Chikashige, R.N. Mohapatra and R.D. Peccei,
\newblock Phys. Rev. Lett. 45 (1980) 1926.

\bibitem{Gelmini:1981re}
G.B. Gelmini and M. Roncadelli,
\newblock Phys. Lett. B99 (1981) 411.

\bibitem{Pakvasa:1999ta}
S. Pakvasa,
\newblock hep-ph/0004077.

\bibitem{Smirnov:1992eg}
A.Y. Smirnov and G.T. Zatsepin,
\newblock Mod. Phys. Lett. A7 (1992) 1272.

\bibitem{Aharonov:1988ju}
Y. Aharonov, F.T. Avignone and S. Nussinov,
\newblock Phys. Lett. B200 (1988) 122.

\bibitem{Giunti:1991ca}
C. Giunti, C.W. Kim and U.W. Lee,
\newblock Phys. Rev. D44 (1991) 3635.

\bibitem{Giunti:1998wq}
C. Giunti and C.W. Kim,
\newblock Phys. Rev. D58 (1998) 017301, hep-ph/9711363.

\bibitem{Grimus:1998uh}
W. Grimus, P. Stockinger and S. Mohanty,
\newblock Phys. Rev. D59 (1999) 013011, hep-ph/9807442.

\bibitem{Cardall:1999ze}
C.Y. Cardall,
\newblock Phys. Rev. D61 (2000) 073006, hep-ph/9909332.

\end{thebibliography}
\end{document}